\PassOptionsToPackage{unicode}{hyperref}
\PassOptionsToPackage{hyphens}{url}
\PassOptionsToPackage{dvipsnames,svgnames,x11names}{xcolor}
\documentclass[
]{article}
\usepackage{amsmath,amssymb}
\usepackage{lmodern}
\usepackage{iftex}
\ifPDFTeX
  \usepackage[T1]{fontenc}
  \usepackage[utf8]{inputenc}
  \usepackage{textcomp} % provide euro and other symbols
\else % if luatex or xetex
  \usepackage{unicode-math}
  \defaultfontfeatures{Scale=MatchLowercase}
  \defaultfontfeatures[\rmfamily]{Ligatures=TeX,Scale=1}
\fi
\IfFileExists{upquote.sty}{\usepackage{upquote}}{}
\IfFileExists{microtype.sty}{% use microtype if available
  \usepackage[]{microtype}
  \UseMicrotypeSet[protrusion]{basicmath} % disable protrusion for tt fonts
}{}
\makeatletter
\@ifundefined{KOMAClassName}{% if non-KOMA class
  \IfFileExists{parskip.sty}{%
    \usepackage{parskip}
  }{% else
    \setlength{\parindent}{0pt}
    \setlength{\parskip}{6pt plus 2pt minus 1pt}}
}{% if KOMA class
  \KOMAoptions{parskip=half}}
\makeatother
\usepackage{xcolor}
\newlength{\cslhangindent}
\newlength{\csllabelwidth}
\newlength{\cslentryspacingunit} % times entry-spacing
\newenvironment{CSLReferences}[2] % #1 hanging-ident, #2 entry spacing
 {% don't indent paragraphs
  \setlength{\parindent}{0pt}
  \ifodd #1
  \let\oldpar\par
  \def\par{\hangindent=\cslhangindent\oldpar}
  \fi
  \setlength{\parskip}{#2\cslentryspacingunit}
 }%
 {}
\usepackage{calc}

\ifLuaTeX
\usepackage[bidi=basic]{babel}
\else
\usepackage[bidi=default]{babel}
\fi
\babelprovide[main,import]{american}
\def\languageshorthands#1{}
\ifLuaTeX
  \usepackage{selnolig}  % disable illegal ligatures
\fi
\IfFileExists{bookmark.sty}{\usepackage{bookmark}}{\usepackage{hyperref}}
\IfFileExists{xurl.sty}{\usepackage{xurl}}{} % add URL line breaks if available
\hypersetup{
  pdftitle={ronswanson: Building Table Models for 3ML},
  pdfauthor={J. Michael Burgess},
  pdflang={en-US},
  colorlinks=true,
  linkcolor={Maroon},
  filecolor={Maroon},
  citecolor={Blue},
  urlcolor={Blue},
  pdfcreator={LaTeX via pandoc}}

\title{ronswanson: Building Table Models for 3ML}

\usepackage[affil-it]{authblk}
\usepackage{orcidlink}
\author[1%
  ]{J. Michael Burgess%
    \,\orcidlink{0000-0003-3345-9515}\,%
    }

\affil[1]{Max Planck Institute for Extraterrestrial Physics,
Giessenbachstrasse, 85748 Garching, Germany}
\date{13 October 2022}

\begin{document}
\maketitle

\hypertarget{summary}{%
\section{Summary}\label{summary}}

\texttt{ronswanson} provides a simple-to-use framework for building
so-called table or template models for \texttt{astromodels}
(\protect\hyperlink{ref-astromodels}{Vianello et al., 2021}) the
modeling package for multi-messenger astrophysical data-analysis
framework, \texttt{3ML} (\protect\hyperlink{ref-threeml}{Vianello et
al., 2015-07}). With \texttt{astromodels} and \texttt{3ML} one can build
the interpolation table of a physical model result of an expensive
computer simulation. This then enables efficient reevaluation of the
model while, for example, fitting it to a dataset. While \texttt{3ML}
and \texttt{astromodels} provide factories for building table models,
the construction of pipelines for models that must be run on
high-performance computing (HPC) systems can be cumbersome.
\texttt{ronswanson} removes this complexity with a simple, reproducible
templating system. Users can easily prototype their pipeline on
multi-core workstations and then switch to a multi-node HPC system.
\texttt{ronswanson} automatically generates the required \texttt{Python}
and \texttt{SLURM} scripts to scale the execution of \texttt{3ML} with
\texttt{astromodel}'s table models on an HPC system.

\hypertarget{statement-of-need}{%
\section{Statement of need}\label{statement-of-need}}

Spatio-spectral fitting of astrophysical data typically requires
iterative evaluation of a complex physical model obtained from a
computationally expensive simulation. In these situations, the
evaluation of the likelihood is computationally intractable even on HPC
systems. To circumvent this issue, one can create a so-called template
or table model by evaluating the simulation on a grid of parameter
values, then interpolating this output. Several spectral fitting
packages, including \texttt{XSPEC}
(\protect\hyperlink{ref-xspec}{Arnaud, 1996}), \texttt{3ML}, and
\texttt{gammapy} (\protect\hyperlink{ref-acero}{Acero et al., 2023};
\protect\hyperlink{ref-gammapy}{Deil et al., 2017}), implement
frameworks that allow for the reading these template models in various
file formats. However, none of these libraries provide tools for
uniformly generating the data from which these templates are built.
\texttt{ronswanson} builds table models for \texttt{astromodels}, the
modeling language of the multi-messenger data analysis framework
\texttt{3ML} in an attempt to solve this problem. \texttt{astromodels}
stores its table models as \texttt{HDF5}
(\protect\hyperlink{ref-hdf5}{The HDF Group, 1997-NNNN}) files. While
\texttt{astromodels} provides a set of user-friendly factories for
constructing table models, the workflow for using these factories on
desktop workstations or HPC systems can be complex. However, these
workflows are easily abstracted to a templating system that can be
user-friendly and reproducible.

\hypertarget{procedure}{%
\section{Procedure}\label{procedure}}

Once the user selects a simulation from which they would like to create
a table model, the first task is to create an interface class that tells
\texttt{ronswanson} how to run the simulation and collect its output.
This is achieved by inheriting a class from the package called
\texttt{Simulation} and defining its virtual \texttt{run} member
function. With this function, the user specifies how the model
parameters for each point in the simulation grid are fed to the
simulation software. The outputs from the simulation are passed to a
dictionary with a key for each different output. Finally, this
dictionary is returned from the \texttt{run} function. This is all the
programming that is required as \texttt{ronswanson} uses this subclass
to run the simulations on the user's specified architecture.

With the interface to the simulation defined, the user must specify the
grid of parameters on which to compute the output of the simulation.
This is achieved by specifying the grid points of each parameter in a
YAML file. Parameter grids can either be custom, or specified with
ranges and a specific number of evaluation points. Additionally, the
energy grid corresponding to the evaluation of each of the simulation
outputs must be specified in this file. The final step is to create a
YAML configuration file telling \texttt{ronswanson} how to create the
table. This includes specifying the name of the output HDF5 database,
where to find the simulation subclass created in the first step, the
name of the parameter YAML file, and details on the compute architecture
on which the simulation grid is to be run.

With these two configuration files defined, the user runs the command
line program \texttt{simulation\_build} on the main configuration file.
This automatically generates all the required \texttt{Python} and
\texttt{SLURM} scripts required for the construction of the table model.
If running on a workstation, the user then executes the
\texttt{run\_simulation.py} script. If, instead, the simulation is run
on an HPC cluster, the user runs \texttt{sbatch\ run\_simulation.sh}. In
the case of running on an HPC system, the final step to build the
database requires running \texttt{sbatch\ gather\_results.sh} which uses
\texttt{MPI} (\protect\hyperlink{ref-mpi}{Forum, 1994}) to gather the
individual pieces of the simulations into the main database.

The created \texttt{HDF5} database can be loaded with utilities in
\texttt{ronswanson} to then construct a table model in the
\texttt{astromodels} format
(\href{https://threeml.readthedocs.io/en/stable/notebooks/spectral_models.html\#Template-(Table)-Models}{see
here for details}). This intermediate step allows the user to select
subsets of the parameters from which to construct the table model. This
is useful as large interpolation tables can consume a lot of computer
memory, and it is possible that certain fits may only need a limited
parameter range. Additionally, utilities are provided that allow adding
parameter sets onto the primary database to extend the interpolation
range. Moreover, the database stores information such as the runtime of
each grid point of the simulation. Utilities are provided to view this
metadata. With future interfacing of \texttt{3ML} and \texttt{gammapy},
these table models could even be used to fit data from optical to very
high energy gamma-rays. More details and examples can be found in the
\href{http://jmichaelburgess.com/ronswanson/index.html}{documentation}.

\hypertarget{acknowledgments}{%
\section{Acknowledgments}\label{acknowledgments}}

This project was inspired by earlier works of Elisa Schoesser and
Francesco Berlato.

\hypertarget{references}{%
\section*{References}\label{references}}
\addcontentsline{toc}{section}{References}

\hypertarget{refs}{}
\begin{CSLReferences}{1}{0}
\leavevmode\vadjust pre{\hypertarget{ref-acero}{}}%
Acero, F., Aguasca-Cabot, A., Buchner, J., Carreto Fidalgo, D., Chen,
A., Chromey, A., Contreras Gonzalez, J. L., Bony de Lavergne, M. de,
Miranda Cardoso, J. V. de, Deil, C., Donath, A., Giunti, L., Hinton, J.,
Jouvin, L., Khelifi, B., King, J., Lefaucheur, J., Lenain, J.-P.,
Linhoff, M., \ldots{} Wood, M. (2023). \emph{Gammapy: Python toolbox for
gamma-ray astronomy} (Version v1.0.1). Zenodo.
\url{https://doi.org/10.5281/zenodo.7734804}

\leavevmode\vadjust pre{\hypertarget{ref-xspec}{}}%
Arnaud, K. A. (1996). {XSPEC: The First Ten Years}. In G. H. Jacoby \&
J. Barnes (Eds.), \emph{Astronomical data analysis software and systems
v} (Vol. 101, p. 17).

\leavevmode\vadjust pre{\hypertarget{ref-gammapy}{}}%
Deil, C., Zanin, R., Lefaucheur, J., Boisson, C., Khelifi, B., Terrier,
R., Wood, M., Mohrmann, L., Chakraborty, N., Watson, J., Lopez-Coto, R.,
Klepser, S., Cerruti, M., Lenain, J. P., Acero, F., Djannati-Atali, A.,
Pita, S., Bosnjak, Z., Trichard, C., Arribas, M. P. (2017).
{Gammapy - A prototype for the CTA science tools}. \emph{35th
International Cosmic Ray Conference (ICRC2017)}, \emph{301}, 766.
\url{https://doi.org/10.22323/1.301.0766}

\leavevmode\vadjust pre{\hypertarget{ref-mpi}{}}%
Forum, M. P. (1994). \emph{MPI: A message-passing interface standard}.
University of Tennessee.

\leavevmode\vadjust pre{\hypertarget{ref-hdf5}{}}%
The HDF Group. (1997-NNNN1997-NNNN). \emph{{Hierarchical Data Format,
version 5}}.

\leavevmode\vadjust pre{\hypertarget{ref-astromodels}{}}%
Vianello, G., Burgess, J. M., Fleischhack, H., Di Lalla, N., \& Omodei,
N. (2021). \emph{Astromodels} (Version 2.2.2). Zenodo.
\url{https://doi.org/10.5281/zenodo.5646925}

\leavevmode\vadjust pre{\hypertarget{ref-threeml}{}}%
Vianello, G., Lauer, R. J., Younk, P., Tibaldo, L., Burgess, J. M.,
Ayala, H., Harding, P., Hui, M., Omodei, N., \& Zhou, H. (2015-07).
\emph{{The Multi-Mission Maximum Likelihood framework (3ML)}}.
\emph{1507}, arXiv:1507.08343. \url{https://doi.org/10.22323/1.312.0130}

\end{CSLReferences}

\end{document}